\documentstyle[12pt]{article}

\newcommand{\be}{\begin{equation}}
\newcommand{\ee}{\end{equation}}
\newcommand{\bea}{\begin{eqnarray*}}
\newcommand{\eea}{\end{eqnarray*}}
\newcommand{\ba}{\begin{eqnarray}}
\newcommand{\ea}{\end{eqnarray}}

\begin{document}
\begin{flushright} UCL-IPT-95-16 \end{flushright}

\begin{center}\large{\bf Symmetry breaking induced by
top quark loops from a model without scalar mass.}\end{center}

\vspace*{15mm}

\begin{center}{T. Hambye}\end{center}

\vspace*{5mm}

\begin{center}
Institut de Physique Th\'eorique\\
UCL \\
 B-1348 Louvain-la-Neuve,
Belgium.
\end{center}

\vspace*{70mm}
\begin{center}
{\bf Abstract} \end{center}
Considering the standard model as an effective electroweak theory,
in which we have no scalar mass term in the Higgs potential
$(\mu^2=0)$, we show that the spontaneous symmetry breaking of
$SU(2)_L
\times U(1)$ can be induced by top loops. The Higgs boson mass
obtained is smaller than 300 GeV.

\vspace*{10mm}
\begin{flushright}09-10-95\end{flushright}
\newpage

In this letter, to implement  the
spontaneous symmetry breaking of $SU(2)_L \times U(1)$, we consider an
effective model, i.e. a model with a
cut-off at some physical scale $\Lambda$. The
lagrangian of this model is the lagrangian of the standard model
with the following Higgs potential
\be
V(\Phi) = \lambda (\Phi^+ \Phi)^2
\ee
where $\Phi$ is the scalar doublet
\be
\Phi = \frac{1}{\sqrt 2} \left(\begin{array}{c}
\phi_2 + i \phi_3 \\ \phi + i \phi_1 \end{array} \right)
\ee
Consequently we assume a scalar quartic interaction $(\lambda \neq
0)$ but no scalar mass term $(\mu^2 = 0)$. The Higgs potential of
Eq.(1), with $\mu^2 =0$, cannot produce the spontaneous symmetry
breaking  (SSB) at tree level. Nevertheless we show that in this
model the one loop effects coming from the top Yukawa force induce
the SSB of
$SU(2)_L \times U(1)$. In this way the SSB induced by top loops is
responsible (and this is the important point) for all the particles
masses, i.e. the  Higgs boson mass (which is expected to be
smaller than 300 GeV) as well as the fermions and gauge bosons
masses.

Before considering the model in detail, it is useful to
situate it in its context. The mechanism of SSB by loop
corrections considered here, which is based on the effective
potential formalism, has been introduced by S. Coleman and E.
Weinberg (CW) in ref.\cite{1}. However the version of the model used
here to implement this mechanism is different from the one
considered by CW in ref.\cite{1}. CW have considered a
renormalizable model where the scalar renormalized mass $\mu_R$ is
put to be 0 in the renormalized Higgs potential (with $\mu^2_R$
defined as
$(\partial^2 V_{eff} (\phi)/\partial \phi^2)|_{\phi=0}$ with
$V_{eff}(\phi)$ the effective potential and $\phi$ the neutral
scalar field). CW have shown that the one loop corrections
coming from the gauge sector (the top Yukawa coupling was neglected
in ref.\cite{1}) induce a SSB. After adjusting $\lambda$ so that
the effective potential has an extremum at $\phi = v \simeq $ 246 GeV
the corresponding square Higgs mass $m^2_H$ (defined by the
expression $(\partial^2 V_{eff} (\phi) / \partial \phi)|_{\phi=v}$)
has been obtained to be $m^2_H = (3 m^4_W + \frac{3}{2}
m^4_Z)/4\pi^2v^2
\simeq$ (9.8 GeV)$^2$. As is well-known this value is excluded by
the present experimental lower limit $m_H$$
{>\hspace{-3.5mm}\raisebox{-1mm}{$\scriptstyle \sim$}}$ 65 GeV (see
for example ref.
\cite{2}).
In addition, when we take in the CW model the top Yukawa coupling
contribution, the corresponding value obtained is $m^2_H = (3m^4_W
+ \frac{3}{2} m^4_Z - 2 N_c m^4_t)/4\pi^2 v^2$ (with $m_t$ the top
quark mass and $N_c$ the number of colors). With the present
experimental value $m_t =$ (180 $\pm$ 12) GeV (see for example ref.
\cite{2} from the experimental results of ref.\cite{3}) we obtain
$m^2_H = -$(50 GeV)$^2$ which means  that the extremum in $\phi=v$
is not a minimum but a maximum and  therefore the CW model does not
work at all.

In the effective model considered in the present letter the top
Yukawa contribution is dominant and large as in the CW
renormalizable model (because $m_t$ is large) but contrary to the
CW model case is responsible for a minimum in $\phi=v$ i.e. for
SSB. Therefore to consider the standard model as an effective theory
(with no mass term in eq.(1)) instead of a renormalizable theory
(with $\mu^2_R = 0$) implies a different situation for SSB. It is
interesting to note  that in the effective theory the
meaning of ``no mass" for the scalar fields before SSB is not the
same as in the renormalizable theory. In the renormalizable theory of
CW there is a quadratic counterterm in
the scalar fields, which means a mass term in the bare lagrangian.
The mass which is put to be zero in the CW renormalizable model is
the renormalized mass which is renormalization scheme dependant. The
choice of the scheme where the renormalized mass is put to be zero
must consequently be justified by a physical argument. In
the effective theory ``no mass" before SSB simply means no quadratic
term in the scalar part of the effective lagrangian, i.e. no
quadratic term coming from the physics at the scale $\Lambda$.

The model considered in this letter has also to be put
in relation with the two models (one renormalizable and one
effective) considered in ref.\cite{4}. The three models are based
on a SSB mechanism due to the top Yukawa interaction. However the
starting assumption of the two models of ref.\cite{4} is different:
in these two models instead of having no mass term $(\mu^2=0)$ and a
quartic term $(\lambda \neq 0)$, there is no quartic term
$(\lambda=0)$ but a mass term $(\mu^2 \neq 0)$.

Let us now consider the model. In the standard model considered as
an effective theory with the
Higgs potential of Eq.(1) the effective potential is
\ba
&&V_{eff}(\phi) = \frac{\lambda}{4} \phi^4 +
\frac{1}{32\pi^2}\int^{\Lambda^2}_0 dq^2q^2 \nonumber\\
&&\cdot
\{A(3\lambda)+3A(\lambda)+6A(\frac{g^2_2}{4})+
3A(\frac{g^2_1+g^2_2}{4})-4N_cA(\frac{g^2_t}{2})\}
 \nonumber\\
&&\hspace{5mm}= \frac{\lambda}{4} \phi^4 + \frac{1}{32\pi^2}.
\nonumber \\
&&\cdot
\{I(3\lambda)+3I(\lambda)+6I(\frac{g^2_2}{4})+3I(
\frac{g^2_1+g^2_2}{4})-4N_c(\frac{g^2_t}{2})\}
\ea
with:
\ba
A(z) &=& \ln (1+\frac{z\phi^2}{q^2})\\
I(z) &=& \frac{1}{2} [\Lambda^4
\ln(1+\frac{z\phi^2}{\Lambda^2})-z^2\phi^4
\ln(1+\frac{\Lambda^2}{z\phi^2})+\Lambda^2z\phi^2].
\ea
In Eq.(3) we have neglected all the Yukawa forces except for the
top quark.
$g_2, g_1, g_t$ are the gauge couplings and the top Yukawa coupling
respectively. In our normalization the $W, Z$ and top masses are
given by $m^2_W = g^2_2 v^2/4,m^2_Z =
(g^2_1+g^2_2)v^2/4,m^2_t=g^2_tv^2/2$ with $v
\simeq $ 246 GeV. In the numerical results the values $m_Z = 91.19$
GeV and $m_W$ = 80.28 GeV will be used.

There is a range of values of $\Lambda$ and $\lambda$ such that
the effective potential has an extremum in $\phi=v$. This range is
defined by the equation:
\ba
0 &=& \frac{\partial V_{eff}(\phi)}{\partial \phi} |_{\phi=v} =
\lambda (v^3+\frac{3v}{8\pi^2}\Lambda^2) \nonumber\\
&+& \frac{1}{8\pi^2v}
\{\frac{1}{2}[-9\lambda^2v^4\ln(1+\frac{\Lambda^2}{3\lambda
v^2})]+\frac{3}{2}[-\lambda^2v^4\ln(1+\frac{\Lambda^2}{\lambda
v^2})]\}\nonumber\\
&+&
\frac{1}{8\pi^2v}
\{-2N_c[m^2_t\Lambda^2-m^4_t\ln(1+\frac{\Lambda^2}{m^2_t})]
\nonumber\\
&&\hspace{10mm} + 3 [m^2_W\Lambda^2 - m^4_W \ln
(1+\frac{\Lambda^2}{m^2_W})] \nonumber \\
&&\hspace{10mm} + \frac{3}{2} [m^2_Z \Lambda^2 - m^4_Z \ln
(1+\frac{\Lambda^2}{m^2_Z})]\}
\ea

The numerical solution of Eq.(6) for $\lambda$ is represented as a
function of $\Lambda$ in Fig.1. From Eq.(6) it is easy to see that
for
$\Lambda
\to
\infty,
\lambda$ goes to the  asymptotical value
\be
\lambda \stackrel{\Lambda \to \infty}{\longrightarrow} (g^2_t -
\frac{1}{4}g^2_2 - \frac{1}{8}(g^2_1+g^2_2))
\ee
For $m_t$ = 180 GeV it corresponds to $\lambda = $ 0,90.

 From the numerical solution of Eq.(6) and from the definition
\ba
m^2_H &=&
\left(\frac{\partial^2V_{eff}(\phi)}{\partial\phi^2}\right)|_{\phi=v}
= \lambda (3v^2+\frac{3}{8\pi^2}\Lambda^2)\nonumber \\
&+&
\frac{1}{8\pi^2v^2}\{\frac{1}{2}[\frac{18\lambda^2v^4}{\Lambda^2+3\lambda
v^2}\Lambda^2-27\lambda^2v^4\ln(1+\frac{\Lambda^2}{3\lambda v^2})]
\nonumber\\
&&+ \frac{3}{2}
[\frac{2\lambda^2v^4}{\Lambda^2+\lambda
v^2}\Lambda^2-3\lambda^2v^4\ln(1+\frac{\Lambda^2}{\lambda v^2})]\}
\nonumber\\
&+ \displaystyle{\frac{1}{8\pi^2v^2}}&\{- 2N_c
[(m^2_t+\frac{2m^4_t}{\Lambda^2+m^2_t})\Lambda^2-3m^4_t\ln(1+
\frac{\Lambda^2}{m^2_t})]
\nonumber\\
&&+3[(m^2_W+\frac{2m^4_W}{\Lambda^2+m^2_W})
\Lambda^2-3m^4_W\ln(1+\frac{\Lambda^2}{m^2_W})]\nonumber\\
&&+\frac{3}{2}[(m^2_Z+\frac{2m^4_Z}{\Lambda^2+m^2_Z})
\Lambda^2-3m^4_Z\ln(1+\frac{\Lambda^2}{m^2_Z})]\}
\ea
we can now obtain the corresponding Higgs mass which is represented
in Fig.2 as a function of $\Lambda$ for $m_t $ = 180 GeV. From
Fig.2 we see there is SSB for 250 GeV
${<\hspace{-3.5mm}\raisebox{-1mm}{$\scriptstyle \sim$}}$ $\Lambda$
${<\hspace{-3.5mm}\raisebox{-1mm}{$\scriptstyle \sim$}}$ 10$^{7.8}$
GeV and the predicted Higgs mass is bigger than the experimental
lower limit, $m_H$ ${>\hspace{-3.5mm}\raisebox{-1mm}{$\scriptstyle
\sim$}}$ 65 GeV, for 350 GeV
${<\hspace{-3.5mm}\raisebox{-1mm}{$\scriptstyle \sim$}}$ $\Lambda$
${<\hspace{-3.5mm}\raisebox{-1mm}{$\scriptstyle \sim$}}$ 10$^{7.6}$
GeV. It can be shown for these ranges of values of $\Lambda$ that
the effective potential is a double well potential bounded from
below. In addition, and this is the important predictive result of
the model, whatever the value of $\Lambda$ is, the Higgs mass is
lighter than a relatively low upper limit: $m_H$
${<\hspace{-3.5mm}\raisebox{-1mm}{$\scriptstyle \sim$}}$ 300 GeV.

An important characteristic of the model presented here is that in
 Eq.(8) there is no quadratic term in $\Lambda$ for
$\Lambda \to \infty$. That can be understood easily from Eq.(7) which
is the condition of cancellation of quadratic divergences in the
ordinary standard model \cite{5}. The quadratic divergences come
from the tadpole diagrams of the scalar field.
The extremum condition (Eq.(6)) is, by definition of the effective
potential, the condition which imposes to the scalar field to have a
zero one-point function and   consequently, in a effective theory,
to have no quadratic divergences in Eq.(8). This cancellation of
quadratic terms in
$\Lambda$ explains the relatively weak dependance of $m^2_H$
on
$\Lambda$.

An other important characteristic of the model is that, for values
of $\Lambda$ sizeably bigger than 1 TeV, the higher orders scalar
contributions (of the order
$\lambda^3,
\lambda^4$, ...) are expected to be big. Indeed if we don't take
into account the
$\lambda^2$ order terms in Eqs.(6), (8), we obtain for
$m_H$ a result which for $\Lambda >>$ 1 TeV
is sizeably different from the result we obtain when we don't take
into account the $\lambda^2$ order terms (see Fig.2). So, for values
of $\Lambda >>$ 1 TeV, the validity of the perturbation
theory is doubtful and the one loop approximation is expected to
break down. In the following we will discuss the results only for
$\Lambda$ ${<\hspace{-3.5mm}\raisebox{-1mm}{$\scriptstyle \sim$}}$
10 TeV\footnote{If we consider as  serious the problem of
fine-tuning, related to the cancellation of quadratic divergences,
let us recall that to take  $\Lambda$ of the order of 1 TeV avoids
this problem.}.

In Fig.3 we plot with better accuracy $m^2_H$ coming from Eqs.(6),
(8), for $\Lambda \leq $ 10 TeV. To give an idea, if $\Lambda$ = 0.5
TeV, 1 TeV, 5 TeV, 10 TeV, we get from Eqs.(6), (8), $m_H$ = 112
GeV, 206 GeV, 297 GeV, 289 GeV. If we don't take into account the
$\lambda^2$ order terms we get from Eqs.(6), (8):
\ba
m^2_H &=&
\frac{1}{8\pi^2v^2}
\{-2N_cm^4_t[\frac{2\Lambda^2}{\Lambda^2+m^2_t}-2\ln(1+\frac{\Lambda^2}{m^2_t})]
\nonumber\\
&& \hspace{12mm} +
3m^4_W[\frac{2\Lambda^2}{\Lambda^2+m^2_W}-2\ln(1+\frac{\Lambda^2}{m^2_W})]
\nonumber\\
&& \hspace{12mm} +
\frac{3}{2}
m^4_Z[\frac{2\Lambda^2}{\Lambda^2+m^2_Z}-2\ln(1+\frac{\Lambda^2}{m^2_Z})]\}
\nonumber\\
&-&
\frac{1}{4\pi^2}\frac{1}{(v^2+\frac{3}{8\pi^2}\Lambda^2)}
\{2N_cm^4_t\ln(1+\frac{\Lambda^2}{m^2_t})
\nonumber\\
&&\hspace{25mm}
- -3m^4_W\ln(1+\frac{\Lambda^2}{m^2_W}) \nonumber\\
&&\hspace{25mm}
- -\frac{3}{2} m^4_Z \ln (1+\frac{\Lambda^2}{m^2_Z})\}\nonumber \\
&+&
\frac{1}{4\pi^2}\frac{\Lambda^2}{(v^2+\frac{3}{8\pi^2}\Lambda^2)}(2N_cm^2_t-
3m^2_W-\frac{3}{2}m^2_Z)
\ea
For $\Lambda$ = 0,5 TeV, 1 TeV, 5 TeV, 10 TeV, Eq.(9) gives $m_H$ =
115 GeV, 206 GeV, 339 GeV, 352 GeV respectively. In Eq.(9), the
third term is dominant: it gives $m_H$ = 121 GeV, 204 GeV, 319 GeV,
326 GeV respectively.\footnote{Note here that for $\Lambda \simeq$
1 TeV the third term of Eq.(9) is an excellent approximation of the
full one-loop result.}
This third term of Eq.(9) is consequently a good approximation
(valid at
$\simeq$
$10\%$) of the complete one loop result for $\Lambda$
${<\hspace{-3.5mm}\raisebox{-1mm}{$\scriptstyle \sim$}}$ 10 TeV.
For $\Lambda^2 < \frac{8\pi^2}{3}v^2(\simeq$(1260GeV)$^2$) and for
$\Lambda^2 > \frac{8\pi^2}{3}v^2$ the third term of Eq.(9) can be
written as
\ba
&&m^2_H \stackrel{\Lambda^2<8\pi^2v^2/3}{\simeq}
\frac{1}{4\pi^2}\frac{\Lambda^2}{v^2}(2N_cm^2_t-3m^2_W-\frac{3}{2}m^2_Z)+{\cal
O}(\frac{3\Lambda^2}{8\pi^2v^2}) \\
&&m^2_H \stackrel{\Lambda^2>8\pi^2v^2/3}{\simeq}
(\frac{4}{3}N_cm^2_t-2m^2_W-m^2_Z)+{\cal O}(\frac{8\pi^2v^2}{3
\Lambda^2})
\\
&&\hspace{8mm} \stackrel{m_t=180GeV}{\simeq } (329GeV)^2 + {\cal
O}(\frac{8\pi^2v^2}{3\Lambda^2})
\ea
Eq.(10) explains why $m_H$ increases quickly in Fig.3 for small
value of $\Lambda$. Eq.(11) explains the plateau in Fig.3 for
$\Lambda$  ${>\hspace{-3.5mm}\raisebox{-1mm}{$\scriptstyle \sim$}}$
2 TeV. Indeed, for $\Lambda^2 > (8\pi^2v^2/3)$, Eq.(11) shows that
up to small corrections (the correction of the order
$8\pi^2v^2/3\Lambda^2$, the two first terms in Eq.(9) and the
$\lambda^2$ order terms which together give for example, for
$\Lambda$ = 3 TeV, a $\simeq$ 10\% correction), $m^2_H$ can be
reduced to the expression of Eq.(11). Neglecting the ${\cal
O}(8\pi^2v^2/3\Lambda^2)$ term, Eq.(11) is the equation obtained by
imposing the cancellation of quadratic divergences in the ordinary
standard model at lowest order
\cite{5}. This result can be understood in the calculation by the
fact that the coefficients of the
$\Lambda^2/4\pi^2(v^2+3\Lambda^2/8\pi^2)$ term  in the dominant third
term of Eq.(9) come
from the coefficients of the quadratically divergent terms in
Eqs.(3)-(5).

Note also that in the third term of Eqs.(9)-(11) we see clearly
that a heavy top is an essential ingredient for the quantum SB
mechanism proposed here. A real Higgs boson requires:
\be
m^2_t {>\hspace{-3.5mm}\raisebox{-1mm}{$\scriptstyle \sim$}}
(\frac{1}{2}m^2_W+\frac{1}{4}m^2_Z) \simeq (73 GeV)^2
\ee
In Eq.(13), the masses are squared because the contribution of each
sector in the dominant third term of Eq.(9) is proportional to the
corresponding squared mass. This explains that the gauge bosons
contribution is a small effect with respect to the top Yukawa
contribution but is not negligible.\footnote{In the models of
ref.\cite{4} the contribution of each sector is proportional to the
4th power of the mass. In this case the gauge bosons contribution
is negligible.} Eq.(9) without the gauge bosons contributions gives
for $\Lambda$ = 0,5 TeV, 1 TeV, 5 TeV, 10 TeV, $m_H$ = 126 GeV, 226
GeV, 368 GeV, 382 GeV respectively, to be compared with $m_H$ = 115
GeV, 206 GeV, 339 GeV, 352 GeV when we take the full Eq.(9).

To conclude, we propose, starting from an effective theory without
a scalar mass term, a dynamical SSB mechanism which is due to the
large top Yukawa force. We obtain the relatively low upper limit:
$m_H {<\hspace{-3.5mm}\raisebox{-1mm}{$\scriptstyle \sim$}} $ 300
GeV. We have $m_H {>\hspace{-3.5mm}\raisebox{-1mm}{$\scriptstyle
\sim$}}$ 100 GeV if $\Lambda
{>\hspace{-3.5mm}\raisebox{-1mm}{$\scriptstyle \sim$}}$ 0.5 TeV and
over a wide range of values of $\Lambda$ (2 TeV
${<\hspace{-3.5mm}\raisebox{-1mm}{$\scriptstyle \sim$}}\Lambda
{<\hspace{-3.5mm}\raisebox{-1mm}{$\scriptstyle \sim$}}$ 10 TeV) we
obtain $m_H \simeq$ 290 GeV.

\vspace{15mm}

\par\noindent {\Large{\bf Acknowledgments}}

We gratefully acknowledge many discussions with J.P. Fatelo, J.-M.
G\'erard, J. Pestieau and J. Weyers. As ``chercheur I.I.S.N.", we
are also indebted to the Institut Interuniversitaire
des Sciences Nucl\'eaires for their financial support.

\newpage
\par\noindent {\Large{\bf Figure captions}}

\begin{tabular}{llll}
Fig.1 & $\lambda$ as a function of $\Lambda$ from Eq.(6) for
$m_t=$180 GeV.
\end{tabular}

\begin{tabular}{llll}
Fig.2 &
$m_H$ as a function of $\Lambda$ from Eqs.(6), (8) for $m_t$ = 180
GeV
\\ &taking into account the $\lambda^2$ order terms (solid line)
and
\\ &without taking into account the $\lambda^2$ order terms (dashed
line).
\end{tabular}

\begin{tabular}{llll}
Fig.3 &
$m_H$ as a function of $\Lambda$ from Eqs.(6), (8) taking into
account
\\ & the $\lambda^2$ order terms for $\Lambda \leq $ 10 TeV and for
$m_t$=168 GeV,
\\ &180 GeV, 192 GeV.
\end{tabular}

\newpage

\small{$m_t$ = 192 GeV}\\
\small{$m_t$ = 192 GeV}\\
\small{$m_t$ = 180 GeV}\\
\small{$m_t$ = 180 GeV}\\
\small{$m_t$ = 168 GeV}\\
\small{$m_t$ = 168 GeV}\\
\small{$m_H$ (GeV)}\\
\small{$m_H$ (GeV)}\\
\small{$m_H$ (GeV)}\\
\small{$m_H$ (GeV)}\\
$\lambda$\\
$\lambda$\\
$\log (\Lambda/1$GeV)\\
$\log (\Lambda/1$GeV)\\
$\log (\Lambda/1$GeV)\\$\log (\Lambda/1$GeV)\\
$\Lambda$ (TeV)\\
$\Lambda$ (TeV)\\
$\Lambda$ (TeV)\\$\Lambda$ (TeV)\\

\end{document}